\documentstyle[12pt,epsfig]{article}

\title{\bf  Implications of percolation of colour strings on
multiplicities, correlations and the transverse momentum}
\author{M.A.Braun\\
High-energy physics department\\
S.Petersburg University, 198904 S.Petersburg, Russia\\
\hspace{3mm}  and C.Pajares
 \\ Departamento
de F\'{\i}sica de Part\'{\i}culas,\\ Universidade de Santiago de
Compostela,\\ 15706-Santiago de Compostela, Spain}
\date{July 1999}
\def\beq{\begin{equation}}
\def\eeq{\end{equation}}
\def\noi{\noindent}
\def\avnu{\langle \nu_{n} \rangle}
\def\avm{\langle M \rangle}

\def\avmf {\langle \mu_F \rangle}
\def\avmff {\langle \mu_F^2 \rangle}
\def\avmb {\langle \mu_B \rangle}
\def\avmfb {\langle \mu_F\mu_B \rangle}
\def\avpt{\langle p_T^2\rangle}
\begin{document}
\maketitle
\medskip
Percolation, multiplicities, $\avpt$ and correlations.
\vspace{1 cm}

\noi{\bf Abstract.}

In the colour string model the impact of string percolation on multiplicities,
$\avpt$ and their long range (forward-backward) correlations is studied.
It is assumed that different string overlaps produce the observed hadrons
independently. The multiplicities are shown to be damped by a simple factor
which follows from the percolation theory. The $\avpt$ rise at the same rate
as multiplicities fall.
A clear signature of the percolation phase transition is found to be a 
behaviour of
the forward-backward correlations for intensive quantities, such as $\avpt$ or
 its
inverse, which can be detected in the relativistic heavy ion collider.

\vspace{3.0cm}
\noi{\Large\bf hep-ph/9907332}\\
\noi{\Large\bf US-FT/15-99}
\newpage

\section{Introduction}
Multiparticle production
at high energies is currently (and successfully) described in terms of
colour strings stretched between the projectile and target [1-6].
Hadronization of these strings produces the observed hadrons. 
Colour strings may be viewed upon as (small) areas in the transverse space
filled with colour field created by the colliding partons. Creation of particles
goes via emission of $q\bar q$ pairs in this field. 
With growing energy and/or
atomic number of colliding particles, the number of strings grows.
Once strings have a certain nonzero dimension in the transverse space they
start to overlap  forming clusters, very much like disks in the
2-dimensional percolation theory. The geometrical behaviour of strings in
the transverse plane then follows that of percolating discs. In particular
at a certain critical string density a macroscopic cluster appears (infinite
in the thermodynamic limit), which marks the percolation phase transition
[7 - 9].

The percolation theory governs the geometrical pattern of the string
clustering.
Its observable implications however  require introduction of some dynamics
to describe
string interaction,
that is, the behavior of a cluster formed by several overlapping strings.

One can study different possibilities.

A most naive attitude is to assume that nothing happens as strings overlap, 
in other words,
they continue to emit particles independently without being affected by their
overlapping neighbours. This is a scenario of non-interacting strings,
which closely
corresponds to original calculations in the colour strings approach, oriented 
at comparatively small energies (and numbers of strings). 
This scenario  however contradicts the idea that strings are areas of the 
transversal 
space filled with colour field and thus with energy, since in the overlapping 
areas the energy should have grown.

In another limiting case one may assume that a cluster of several overlapping 
strings
behaves as a single string with an appropriately higher colour field (a string 
of higher colour, or a "colour rope" [10]). This fusion scenario was proposed
by the authors
and later realized as a Monte-Carlo algorithm nearly a decade ago [11,12]. It 
predicts
lowering of total multiplicities and forward-backward correlations (FBC)  
and also
strange baryon enhancement, in a reasonable agreement with the  known 
experimental trends.

However both discussed scenarios are obviously of a limiting sort. In a typical
situation
strings only partially overlap and there is no reason to expect them to fuse 
into a single
stringy object, especially if the overlap is small. The
transversal space occupied by a cluster of overlapping strings splits into
a number of areas
in which different number of string overlap, including areas where no
overlapping takes place.
In each such area colour fields coming from the overlapping strings will add 
together.
As a result the total cluster area is split in domains with different colour 
field strength.
As a first approximation, neglecting the interaction at the domain frontiers, 
one may assume
that emission of $q\bar q$ pairs in the domains proceeds independently, 
governed by the field strength ("the string tension") in a given domain.
This picture implies that  clustering  of strings actually leads to
their proliferation, rather than fusion,
since each particular overlap may be considered as a separate string.
Evidently newly formed strings differ not only in their
colours but also in their transverse areas. 

As a simple example consider a cluster of two partially overlapping strings 
(Fig. 1).
One distinguishes three different regions: regions 1 and 2 where no 
overlapping takes
place and the colour field remains the same as in a single string, and 
the overlap region 3 with colour fields of both strings summed. In our picture 
particle
production will proceed independently from these three areas, that is, from 
three different
"strings" corresponding to areas 1,2 and 3. In this sense string interaction 
has split two
strings into three of different colour, area and form in the
transverse space.

We stress that these dynamical assumptions are  rather independent of the 
geometrical
picture of clusterization. In particular, in each of the scenarios discussed 
above, at a 
certain string density there occurs the percolation phase
transition. However its experimental signatures crucially depend on the
dynamical contents of 
string interaction. With no interaction, clustering does not change physical 
observables,
so that the geometric percolation will not be felt at all. With the 
interaction between strings
turned on, clustering (and percolation) lead to well observable implications.

In this note we shall study these implications for simplest observables, 
such as multiplicities,
average transverse momenta and FBC in the realistic scenario discussed
above, which corresponds
to independent particle production from different overlap domains. Our choice 
of observables
is dictated by the possibility to relate them directly to the domain 
properties, without
introducing any more assumptions.

\section{Multiplicity and $\avpt$ for overlapping strings}
As stated in the Introduction, the central dynamical problem is to find
how the observables change when several strings form a cluster partially
overlapping .
In the overlap areas the colour fields of individual strings are summed 
together.
It is more convenient to sum the charges which generate the colour field of
overlapping strings. 

Let only two strings, of areas $\sigma_0$ in the transverse space each, 
partially overlap in the area $S_2$ (region 3
in Fig. 1), so that $S_1=\sigma_0-S_2$ is the area in each string not
overlapping with the
other. In the following it will be called the overlap area of one string.
A natural assumption seems to be that the average colour density $\xi$ of
the string
in the transverse plane is a constant 
\beq
\xi=Q_0/\sigma_0,
\eeq
where $Q_0$ is a color of the string. For partially overlapping strings 
the colour in
each of the two non-overlapping areas will then be
\beq Q_1=\xi S_1=Q_0(S_1/\sigma_0).
\eeq
The colour in the overlap area $Q_2$ will be a vector sum of the two 
overlapping colours
$\xi S_2$. In this summation the total colour squared should be conserved
[10].
Indeed, $Q_2^2=({\bf Q}_{ov}+{\bf Q}'_{ov})^2$ where ${\bf Q}_{ov}$ and
${\bf Q}'_{ov}$ are the
two vector colours in the overlap area. Since the colours in the two strings 
may generally
be oriented in an arbitrary manner respective to one another, the average  of 
${\bf Q}_{ov}{\bf Q}'_{ov}$ is zero. Then $Q_2^2=Q^2_{ov}+{Q'_{ov}}^2$, which 
leads to
\beq
Q_2=\sqrt{2}\,\xi S_2=\sqrt{2}\,Q_0(S_2/\sigma_0).
\eeq
One observes that, due to its vector nature, the colour in the overlap
is less than the
sum of the two overlapping colours. This phenomenon was first mentioned in [10] 
for the
so-called colour ropes.

Two simplest observables, the multiplicity $\mu$ and the average
transverse momentum
squared $\avpt$ are directly related to the field strength in the string and 
thus to
its generating colour. In fact they are both proportional to the colour [10,13].
Thus, assuming independent emission from the three regions 1,2 and 3 in Fig. 1
we get for the multiplicity
\beq
\mu/\mu_0=2(S_1/\sigma_0)+\sqrt{2}\,(S_2/\sigma_0),
\eeq
where $\mu_0$ is a multiplicity for a single string. 
To find $\avpt$ one has to divide the total transverse momentum squared of
all observed particles by the total multiplicity. In this way for our cluster 
of two
strings we obtain
\beq
\avpt/\avpt_0=\frac{2(S_1/\sigma_0)+\sqrt{2}\sqrt{2}\,(S_2/\sigma_0)}
{2(S_1/\sigma_0)+\sqrt{2}\,(S_2/\sigma_0)}=\frac{2}{2(S_1/\sigma_0)+\sqrt{2}\,
(S_2/\sigma_0)},
\eeq
where $\avpt_0$ is the average transverse momentum squared for  a single
string and we have
used the evident property $2S_1+2S_2=2\sigma_0$ in the second equality.

Generalizing to any number $N$ of overlapping strings we find the total 
multiplicity as
\beq
\mu/\mu_0=\sum_i\sqrt{n_i}\,(S^{(i)}/\sigma_0),
\eeq
where the sum goes over all individual overlaps $i$ of $n_i$ strings having 
areas $S^{(i)}$.
Similarly for the $\avpt$ we obtain
\beq
\avpt/\avpt_0=\frac{\sum_i n_i\,(S^{(i)}/\sigma_0)}{\sum_i\sqrt{n_i}\,
(S^{(i)}/\sigma_0)}=
\frac{N}{\sum_i\sqrt{n_i}\,(S^{(i)}/\sigma_0)}.
\eeq
In the second equality we again used an evident identity 
$\sum_i n_i\,S^{(i)}=N\sigma_0$.
Note that (6) and (7) imply a simple relation between the multiplicity and
transverse 
momentum
\beq
\frac{\mu}{\mu_0}\frac{\avpt}{\avpt_0}=N,
\eeq
which evidently has a meaning of conservation of the total transverse momentum
produced.

Eqs. (6) and (7) do not look  easy to apply. To calculate the
sums over $i$ one
seems to have to identify all individual overlaps of any number of strings 
with their
areas. For a large number of strings the latter may have very complicated 
forms and
their analysis presents great calculational difficulties. However one 
immediately
recognizes that such individual tracking of overlaps is not at all necessary. 
One can
combine  all terms with a given number of overlapping strings $n_i=n$
into a single term, which sums all such overlaps into a total area of exactly 
$n$
overlapping strings $S_n$. Then one finds instead of (6) and (7)
\beq
\mu/\mu_0=\sum_{n=1}^N\sqrt{n}\,(S_n/\sigma_0)
\eeq
and
\beq
\avpt/\avpt_0=
\frac{N}{\sum_{n=1}^N\sqrt{n}\,(S_n/\sigma_0)}.
\eeq

In contrast to individual overlap areas $S^{(i)}$ the total ones $S_n$ can be 
easily
calculated (see Appendix). Let the projections of the strings onto the
transverse space be distributed uniformly in the interaction
area $S$ with a density $\rho$. Introduce a dimensionless parameter
\beq \eta=\rho \sigma_0=N \sigma_0/S.
\eeq
The "thermodynamic limit" corresponds to taking the number of the strings
$N\rightarrow\infty $ keeping $\eta$ fixed. In this limit one readily finds
that the distribution of the overlaps of $n$ strings is Poissonian with a 
mean value
$\eta$:
\beq
p_n=\frac{S_n}{S}=\frac{\eta^n}{n!}e^{-\eta}.
\eeq
From (9) we then find that the multiplicity is damped due to overlapping by a 
factor
\beq
F(\eta)=\frac{\mu}{N\mu_0}=\frac{\langle\sqrt{n}\rangle}{\eta},
\eeq
where the average is taken over the Poissonian distribution (12).

The behaviour of $F(\eta)$ is shown in Fig. 2. It smoothly goes down from 
unity at
$\eta=0$ to values around 0.5 at $\eta=4$ falling as $1/\sqrt{\eta}$ for larger
$\eta$'s. According to (10) the inverse of $F$ shows the
rise of the $\avpt$.
Note that a crude estimate of $F(\eta)$ can be done from the overall 
compression of 
the string area due to overlapping. The fraction of the total area occupied 
by the
strings according to (12) (see also [14]) is given by
\beq
\sum_{n=1}p_n=1-e^{-\eta}.
\eeq
The compression is given by (14) divided by $\eta$. According to our picture the
multiplicity is damped by the square root of the compression factor, so that 
the damping factor is
\beq
F(\eta)=\sqrt{\frac{1-e^{-\eta}}{\eta}}.
\eeq
For all the seeming crudeness of this estimate, (15) is very close to the 
exact result,
as shown in Fig. 2 by a dashed curve.

\section{Percolation}

Percolation is a purely classical mechanism.
Overlapping strings form clusters.
At some critical value of the parameter $\eta$ a phase transition of the 2nd
 order occurs: a cluster appears which
extends over the whole surface (an infinite cluster
in the thermodynamic limit).  
The critical value of $\eta$  is found to be 
$\eta_c\simeq 1.12-1.20$ [15].
Below the phase transition point, for $\eta<\eta_c$, there is no infinite
cluster,
Above the transition point, at $\eta>\eta_c$ an infinite cluster appears
with a probability
\beq
P_{\infty}=\theta (\eta-\eta_c) (\eta-\eta_c)^{\beta}.
\eeq
The critical exponent $\beta$ can be calculated from Monte-Carlo
simulations.
However the universality of critical behaviour, that is, its independence of
the percolating substrate, allows to borrow its value from lattice
percolation, where  
 $\beta=5/36$.

Cluster configuration can be characterized by the
occupation numbers $\avnu$, that is average numbers of clusters made of $n$ 
strings.
 Their behaviour at all values of $\eta$ and $n$
is not known. From scaling considerations in the vicinity of the phase
transition  it has been found [24]
\beq
\avnu= n^{-\tau}F(n^{\sigma} (\eta-\eta_c)),\ \ |\eta-\eta_c|<<1,\ n>>1,
\eeq
where $\tau=187/91$ and $\sigma=36/91$ and the function $F(z)$ is finite
at $z=0$ and falls off exponentially for $|z|\rightarrow\infty$.
Eq. (17) is of limited value, since near $\eta=\eta_c$ the bulk of the
contribution is still supplied by low values of $n$, for which (14) is not
valid. However from (17) one can find non-analytic parts of other
quantities of interest
at the transition point. In particular, one finds a singular part of
the total number of clusters $M=\sum\nu_n$ as
$\Delta\avm=c|\eta-\eta_c|^{8/3}$.
This singularity is quite weak: not only $\avm$ itself but also its two
first derivatives in $\eta$ stay continuous at $\eta=\eta_c$ and only the
third blows up as $|\eta-\eta_c|^{-1/3}$. So one should not expect that
the percolation phase transition will be clearly reflected in some
peculiar behaviour of standard observables.

Indeed we observed that neither the total multiplicity nor $\avpt$ show any
irregularity in the vicinity of the phase transition, that is, at $\eta$ 
around unity.
This is not surprizing since both quantities reflect the overlap structure 
rather
than the cluster one. The connectedness property implied in the latter has no
 effect
on these global observables.

It is remarkable, however, that the fluctuations of these observables carry some
information about the phase transition. The dispersion of the multiplicity
due to overlapping and clustering
can easily be calculated in the thermodynamic limit (see Appendix). The result
 is
shown in Fig. 3. The dispersion shows a clear maximum around $\eta=1$ (in fact
 at
$\eta\simeq 0.7$). So some information of the percolation phenomenon is passed
 to the
total multiplicity, in spite of the fact that it basically does not feel the
connectedness properties of the formed clusters. Of course, due to relation
 (10), the
dispersion of $\avpt$ has a similar behaviour.

We have to warn against a simplistic interpretation of this result.
 The dispersion
shown in Fig. 3 is only part of the total one, which besides includes 
contributions
from the fluctuations inside the strings and also in their number. Below we
 shall
discuss  the relevance and magnitude of these extra contributions.

An intriguing question is a relation between the percolation and formation of
 the
quark-gluon plasma. Formally these phenomena are different. Percolation is
 related to
the connectedness property of the strings. The (cold) quark-gluon plasma 
formation is
related to the density of the produced particles (or, equivalently, the 
density of
their transverse energy ). However in practice percolation and plasma
formation go together. In fact, the transverse energy density inside a single 
string
seems to be sufficient for the plasma formation. Percolation makes the total 
area
occupied by strings comparable to the total interaction area, thus, creating, 
a sizable
area with energy densities above the plasma formation threshold. 

Let us make some crude
estimates. Comparison with the observed multiplicity densities in 
$pp(\bar{p})$ collisons
at present energies fix the number of produced (charged) particles 
per string per unit
rapidity at approximately unity. Taking the average energy of each 
particle as 0.4 GeV
(which is certainly a lower bound), formation length  in the Bjorken 
formula [16]
as 1 fm and the string transverse radius as 0.2 fm [7] we get the 3-dimensional
transverse energy density inside the string as $\sim 3 GeV/{\rm fm}^3$. 
The plasma
threshold is currently estimated to be at $1 GeV/{\rm fm}^3$. So 
it is tempting to say
that the plasma already exists inside strings. This however has little 
physical sense
because of a very small area occupied by a string. One can speak of a 
plasma only when
the total area occupied by a cluster of strings reaches a sizable fraction 
of the total
interaction area. In Fig. 4 we show this fraction for a maximal cluster as a 
function of
$\eta $ calculated by Monte -Carlo simulations in a system of 50 strings.
It grows wit $\eta$ and the fastest growth occurs precisely in the region of the
percolation phase transition: as $\eta$ grows from 0.8 to 1.2  the fraction 
grows from
0.3 to 0.6. With a string cluster occupying more than half of the interacting 
area,
one can safely speak of a plasma formed in that area. 

\section{Dispersions and forward-backward correlations}
As shown in the previous section a definite signal for the percolation 
phase transition
comes from the multiplicity (or $\avpt$) fluctuations due to string clustering.
However these fluctuations are not the only ones. An important contribution 
also comes
from the fluctuations of the multiplicity inside the strings, or  rather 
inside the
individual overlaps of strings, which in our picture play the role of 
independent
particle emitters. To find the total dispersion it is convenient to use a 
formalism of
the generating functions.
Let the total probability to observe $n$  produced particles be ${\cal P}(n)$.
 In our
picture it is given by a convolution of the probability for a given overlap
configuration $P(C)$ with the probabilties for particle production from all
individual overlaps:
\beq
{\cal P}(n)=\sum_CP(C)\sum_{n_1,...n_M}p_1(n_1)...p_M(n_M)\delta_{n,\sum n_i}.
\eeq
Here $C$ is a "configuration", characterized by the total number of the 
overlaps $M$
and their individual properties: area and number of overlapping strings. 
We pass to generating functions
\beq
\Phi(z)=\sum _nz^n{\cal P}(n),\ \ \phi_i(z)=\sum_n z^n p_i(n),
\eeq
for which (18) transforms into
\beq
\Phi(z)=\sum_CP(C)\prod_{i}\phi_i(z).
\eeq
The overall averages (with the probability ${\cal P}$) are given by
\beq
\langle n\rangle=\Phi'(z)_{z=1},\ \ \ \langle n(n-1)\rangle=\Phi''(z)_{z=1}.
\eeq
The averages inside the $i$-th overlap are likewise given by
\beq
 \overline{n_i}=\phi'(z)_{z=1},\ \ \ \overline{n_i(n_i-1)}=\phi''(z)_{z=1}.
\eeq
Using this formulas one readily finds the total dispersion of multiplicity 
as a sum of
two terms
\beq
D^2=D^2_C+D^2_{in}.
\eeq
Here $D^2_C$ is a dispersion due to fluctuation in configurations
\beq
D_C^2=\sum_CP(C)\sum_{ik}\overline{n_i}\
\overline{n_k}-(\sum_CP(C)\sum_i\overline{n_i})^2.
\eeq
It is calculatied according to the distribution ${\cal P}$ assuming that the
numbers of particles produced in individual overlaps are fixed to be 
their averages.
It is this part of the dispesion, which is calculated in the Appendix and 
presented in
Fig.3. The part $D^2_{in}$ is a part of the dispersion due to fluctuations 
inside
the individual overlaps:
\beq
D^2_{in}=\sum_CP(C)\sum_i(\overline{n^2_i}-(\overline{n_i})^2).
\eeq
Its calculation is extremely difficult even in by Monte Carlo simulations, since
it requires identification of all individual overlaps and knowledge of their 
areas.
To simplify we assume that the distribution of produced particles from any 
overlap is
Poissonian, so that $\overline{n^2_i}-(\overline{n_i})^2=\overline{n_i}$. 
Then we
evidently find
\beq
D^2_{in}=\mu.
\eeq

From the experimental data we estimate $\mu_0\simeq 1.1$  for
unit rapidity interval.
Comparison of Fig.2 and Fig. 3 then shows  that for unit rapidity interval
the internal dispersion $D^2_{in}$ is roughly 40 times greater
than the "percolation dispersion" $D^2_C$ at $\eta=0.7$ corresponding to 
the maximum
for the latter. At other $\eta$ the ratio $D^2_{in}/D^2_C$
is still greater. With the growth of the rapidity interval this ratio
diminishes proportionally. However it is obvious  that one
should not expect to clearly see the percolation effects directly in the 
observed
multiplicities.

A better signal for the percolation comes from the FBC, which, as we shall see,
distinguish between the percolation and intrinsic dispersions 
(in fact they depend on their ratio).
The FBC are described by the dependence of the average multiplicity in the
backward hemisphere $\avmb$ on the event multiplicity in the forward
hemisphere $\mu_F$. The data can be fitted by a linear expression [1]
\beq
\avmb=a+b\mu_F,
\eeq
where $a$ and $b$ are given by expectation values
\beq
a=\frac{\avmb\avmff -\avmfb\avmf}{\avmff-\avmf^2},
\eeq
\beq
b=\frac{\langle\mu_F \mu_B\rangle
-\langle\mu_F\rangle\langle\mu_B\rangle}{\langle\mu_F^2\rangle
-\langle\mu_F\rangle^2}.
\eeq
In absence of FBC $\avmfb=\avmf\avmb$ and one obtains $a=\avmf$ and $b=0$.
So the strength of the correlations is given by the coefficient $b$.

To calculate the necessary averages we introduce the probability ${\cal P}(F,B)$
to produce $F(B)$ particles in the forward (backward) hemispheres.
Similar to (18) it is given by a convolution
\beq
{\cal P}(F,B)=\sum_CP(C)\sum_{F_i,B_i}\prod_ip_i(F_i,B_i)\delta_{F,\sum F_i}
\delta_{B,\sum B_i},
\eeq
which transforms into a relation between the generating functions
\beq \Phi(z_Fz_B)=\sum_CP(C)\prod_i\phi(z_F,z_B).\eeq

Instead of (21) we now have
\beq
\langle F\rangle=\left(\frac{\partial\Phi}{\partial z_F}\right)_{z_F=z_B=1}, \ 
\langle F(F-1)\rangle=\left(\frac{\partial^2\Phi}{\partial z_F^2}
\right)_{z_F=z_B=1},\  
\langle FB\rangle=\left(\frac{\partial^2\Phi}{\partial z_F\partial
z_B}\right)_{z_F=z_B=1},
\eeq
formulas similar to the first two for $B$, and similar formulas for the 
averages
over the distributions $p_i$ inside the overlap $i$ with the generating function
$\phi_i(z_F,z_B)$.
Using this formulas we find that the dispersions again split into the 
percolacion (C)
and internal (in) parts.
\beq
D_F^2=D_{F,C}^2+D^2_{F,in},\ \ D^2_{FB}\equiv\langle FB\rangle-\langle F\rangle
\langle B\rangle=D^2_{FB,C}+D^2_{FB,in}.
\eeq
They are given by
\beq
D_{F,C}^2=\sum_CP(C)\sum_{i,k}\overline{F_i}\ \overline{F_k}-
(\sum_{C}P(C)\sum_i\overline{F_i})^2,
\eeq
\beq
D_{FB,C}^2=\sum_CP(C)\sum_{i,k}\overline{F_i}\ \overline{B_k}-
\sum_{C}P(C)\sum_i\overline{F_i}
\sum_{C}P(C)\sum_i\overline{B_i},
\eeq
\beq
D_{F,in}^2=\sum_CP(C)\sum_i(\overline{F_i^2}-(\overline{F_i})^2),
\eeq
\beq
D_{FB,in}^2=\sum_CP(C)\sum_i(\overline{F_iB_i}-\overline{F_i}\ \overline{B_i}).
\eeq

The usual assumption is that there are no FB correlations for particle 
production from
a single emitter (overap $i$ in our case). Then $D_{FB,in}^2=0$. Also from
$n=F+B$ one can relate these dispersions with the overall ones $D_C^2$ and 
$D_{in}^2$:
\beq
D_{F,C}^2=D_{FB,C}^2=(1/4)D_C^2,\ \ D_{F,in}^2=(1/2)D_{in}^2.
\eeq
From (29) one then finds 
\beq
\frac{1}{b}=1+2\frac{D^2_{in}}{D^2_C}.
\eeq
So, the FBC parameter $b$ indeed measures the ratio of internal to percolation
dispersions squared. If we assume that the distribution of particles produced
from an individual overlap is Poissonian then (39) transforms into
\beq
\frac{1}{b}=1+2\frac{\mu}{D^2_C}.
\eeq

Both $\mu$ and $D^2_C$ grow linearly with the number of strings $N$, so in the
thermodynamic limit $b$ does not depend on $N$. However it depends on the 
multiplicity
of a single string $\mu_0$, since $\mu$ is proportional to $\mu_0$ and
$D^2_C$ is proportional to $\mu_0^2$. As mentioned, from the experimental 
data we
find that $\mu_0\simeq 1.1\,y$ where $y$ is the rapidity interval of the 
produced
particles. So the right-hand side of (40) falls with the growth of the 
rapidity window.
It also has a minimum at $\eta\sim 0.7$ corresponding to the maximum $D^2_C$. 
From this
we see that the FBC parameter $b$ also has a maximum near $\eta\sim 0.7$ whose
magnitude grows with the rapidity interval. In Fig. 5 we show the behaviour of
 $b$ as a
function of $\eta$ for two rapidity intervals $y=5$ and $y=9$

One can also study the FBC for $\avpt$. Due to relation (8), for fixed $N$ 
they are
uniquely determined by the fluctuations in $\mu$. It is convenient to choose the
inverse $\avpt$ as an observable:
\beq
\tau=\frac{\mu_0\avpt_0}{\avpt}=\frac{\mu}{N}.
\eeq
Then it is obvious that the parameter $b$ for $\tau$ is the same as for the
multiplicity,
since both percolation and internal dispersions for $\tau$ are obtained 
from those for
$\mu$, divided by $N$. However this equivalence holds only for fixed number 
of strings
$N$. In the realistic situation $N$ fluctuates. The multiplicity is an extensive
observable and carries  these fluctuations directly. They result large and 
smear out 
nearly all traces of percolation. In contrast, $\tau$ is an intensive 
observable and
so feels the fluctuation in $N$ only indirectly, through the fluctuations in the
parameter $\eta$. As a result, the percolation effects for the FBC in $\tau$ are
clearly visible, as we shall see in the next section.

\section{Realistic hadronic and nuclear collisions}
For realistic hadronic and nuclear collisions the total number of strings $N$
grows with energy and fluctuates.
For pp collisions the initial number of colour strings $N$ created at
a given c.m. energy $\sqrt{s}$ can be taken from the well-known
calculations of [1,2]. For pA and AB collisions this number should by
multiplied by the effective number of collisions $C$.
At asymptotic energies for  minimum bias hA and AB collisions the AGK rules 
predict 
\beq
C_A=A\sigma/\sigma_A,\ \ C_{AB}=AB\sigma/\sigma_{AB},
\eeq
where $\sigma\ (\sigma_A,\ \sigma_{AB})$ is the inelastic pp (pA, AB) 
cross-section. 
For hA 
collisions at fixed impact parameter $b$
\beq
 C_{A}=A\sigma T_{A}(b)/(1-\exp (-A \sigma T_{AB})),
\eeq
where $T_A$ is the standard nuclear
profile function (normalized to unity).
The same relation holds for AB collisions in the optical approximation with
\beq
A\rightarrow AB,\ \ T_A\rightarrow T_{AB}(b)=\int d^2c\,T_{A}(c)T_{B}(b-c).
\eeq

The transverse interaction area for pp and hA collsions obviously is of the
order $\sigma$.
For AB collisions it depends on the geometry of the collisions. We restrict
ourselves to central ($b=0$) collisions of identical nuclei, when evidently
the interaction area is $\sigma_{AA}$. The parameter $\eta$ is then calculated
 according
to (11). We have taken the string area
$\sigma_0$ as
\beq
        \sigma_0=  \pi a^2,\ \ a=0.2\ {\rm fm}.
\eeq
in accordance with arguments presented in
[7].
To relate the multiplicity to the experimental data we normalized $\mu$
to the observable central (charged) multiplicity  per unit rapidity in
$pp$ collisions at low energies. This fixes $\mu_0$ for unit rapidity to be
1.1
 The found values of
$ \mu\equiv (dn^{ch}/dy)_{y=0}$  are presented in the Table
 for central S-S and Pb-PB collisions (4d column) together with the 
corresponding
values of c.m.energy $\sqrt{s}$,
$\eta$ and number of strings $N$ (1st to 3d columns, respectively).
To compare, we present the values of $\mu$ found without fusion and percolation 
(independent strings picture) in the 7th column.
One should have in mind that the asymptotic formulas for the number of 
collisions
 (42) - (44) used
to determine the number of strings are not valid at comparatively low 
energies due to
restrictions imposed by energy conservation. As stated in [1], at lower energies
the number of collisions can be roughly obtained multiplying (42) -  (44) by a 
factor
$1/2$. Correspondingly, our values for the multiplicity at two lower energies
 have to be
corrected for energy conservation by a factor of this order. We preferred not to
make this correction, since, in any case, it cannot be determined with any
 degree of
rigour, and the effects we are considering are essential only at high enough
 energies,
where (42)- (44) are hopefully valid.

As we observe, percolation considerably damps the multiplicity at high 
energies: 
at $\sqrt{s}=7000$ GeV for central Pb-Pb collision  from nearly  8000  down to
approximately 3000. This effect was predicted in our earlier papers on the 
string
fusion [11,12] and is now reproduced in various models [17].

As to the FBC, we studied them both for the multiplicity  and the inverse 
$\avpt$
(in fact for the observable $\tau$, defined in (41)). We have taken one half
 of the
total rapidity available for the relevant rapidity window.
The "external" dispersion in the denominator of (39) has to include
also the fluctuations in the number of strings. We have assumed the overall 
distribution
in $N$ to be Poissonian so that $\Delta N/N=1/\sqrt{N}$. As mentioned, for the
multiplicity, due to its extensive character, the dispersion in $N$ simply 
adds to
$D_C$. It is large and absolutely dominates all other contributions. So it is no
surprise that the coefficient $b$ for the multiplicity with percolation (5th
 columns in
the Table) only slightly differs from the one without percolation (8th column).

In contrast, for $\tau$ the fluctuations in $N$ only enter via the dependence
of $\eta$ on $N$. They result smalller than the percolation contribution at the
maximum of the latter but fall with $\eta$ more slowly, so that they dominate at
large $\eta$. As a result, the characteristic peaked form of the parameter $b$
 found at
fixed $N$ (Fig. 5) disppears and $b$ results steadily growing with $\eta$ (6th
 column in
the Tables). However one should note that in the independent string picture 
the parameter
$b$ for $\tau$ is zero, since the dependence on $N$ is then completely absent.
So experimental study of the FBC for inverse $\avpt$ seems to be a promising 
way to 
observe signatures of string fusion and percolation.

\section{Conclusions}
In this study  we have analyzed the impact of fusion and percolation of 
colour strings
on global observables, such as multiplicities and $\avpt$. To do this, a certain
dynamical assumption has been made. The strings have been assumed to decay 
into the
observed hadrons independently in each overlap.

On the qualitative level the results are best seen in the idealized case of a 
fixed
number of strings $N>>1$ (equivalent to plasma studies in the thermodynamic 
limit).
A clear consequence of fusion and percolation is damping of the 
multiplicities, well
described by a damping factor (15), which follows from the percolation theory.
An unexpected but potentially important result is that the parameter $b$ of 
the FBC
shows a clear maximum at the percolation point. Such a maximum would be natural
in  fluctuations, say, of the cluster sizes, where it is to be expected as
a signature of the percolation phase transition. However multiplicities do 
not seem to
feel directly the cluster structure, so that the appearance
of the maximum in their fluctuations is a new result.

In the realistic case of nuclear collisions, where $N$ fluctuates, the predicted
multiplicities repeat the pattern observed at fixed $N$ and are damped by 
the same
factor (15). At the LHC energies it reduces the multiplicities by more than 
two times.
However the parameter $b$ for multiplicity results completely dominated by 
fluctuations
in the number of strings and so only slightly different from the one in the 
independent
string picture. A better (intensive) observable to see the impact of 
percolation is
$\avpt$ or its inverse, for which the contribution of the fluctuations
in $N$ is drastically reduced. For $\avpt^{-1}$ we predict sizable values of 
$b$ to be
contrasted  with $b=0$ in the independent string picture. Observation of 
nonzero values
of $b$ for $\avpt^{-1}$ would therefore be a clear signature of string 
fusion and
percolation.

\section{Acknowledgments}  
M.B. thanks the University of Santiago de Compostela for 
hospitality. C.P. thanks the Institute for Nuclear Theory  of the University of
Washington for hospitality.
This study was supported by the NATO grant CRG.971461.

\section{Appendix. Multiplicities and their dispersion}
Let us recall the geometry of our percolation picture. Discs of radius $a$ 
and area $\sigma_0=\pi a^2$ are homogeneously distributed in the
total area $S$. It is assumed that centers of the discs are inside the unit 
circle
of area $S_0=\pi$ so that $S=\pi (1+a)^2$. The disc density is
$\rho=N/S$ and the percolation parameter is $\eta=\rho\sigma_0=
N\sigma_0/S$. In the thermodynamic limit $N\rightarrow\infty$, so that at fixed
$\eta$ the radius of the discs goes to zero.
For fixed $\eta$
\beq
a=\left(\sqrt{\frac{N}{\eta}}-1\right)^{-1},
\eeq
so that at large $N$ $a\sim 1/\sqrt{N}$ and $\sigma_0\sim1/N$.
Since the discs are distributed homogeneously, the probability that their 
centers are at points $r_i$, $i=1,...N$ inside the unit circle is
independent of $r_i$ and  is  given by
\beq
P(r_i)=S_0^{-N}.
\eeq

Let us take a configuration which corresponds to the disc centers at points 
$r_i$.
Then the overlap area of exactly $n$ discs is given by the integral
\beq
S_n(r_1,...r_N)=\int_Sd^2r\sum_{\{i_1,..i_n\}\subset\{i_1,...i_N\}}
\prod_{k=1}^{n}
\theta(a-|{\bf r}-{\bf r}_{i_k}|)
\prod_{k=n+1}^N\theta(|{\bf r}-{\bf r}_{i_k}|-a).
\eeq
The average of $S_N$ will be given by a multiple integral over $r_i$ with the
probability (47):
\beq
\langle S_n\rangle=\frac{1}{S_0^N}\int_{S_0}\prod_{i=1}^N d^2r_iS_n(r_1,...r_N)=
C_N^n\int_Sd^2rF^n(r)(1-F(r))^{N-n},
\eeq
where
\beq
F(r)=(1/S_0)\int_{S_0}d^2r_1\theta(a-|{\bf r}-{\bf r}_1|).
\eeq

The function $S_0
F(r)$ gives an area occupied by a circle $C$ of radius $a$ with a center
at
$r$ which is inside the unit circle $S_0$. If $r<1-a$ then $C$
is always inside $S_0$ so that
\beq
F(r)=\sigma_0/S_0,\ \ \ 0<r<1-a.
\eeq
However for $r>1-a$ a part of $C$ turns out to be outside the unit circle,
and
\beq
F(r)=\sigma(r)/S_0,\ \ \ 1-a<r<1+a,
\eeq
where $\sigma(r)\leq \sigma_0$ is the overlap of the two discs $C$ and $S_0$.

Generally, an overlap of two circles of radii $r_1$ and $r_2$ with a 
distance $r$
between their centers is given by
\beq
\sigma(r_1,r_2,r)=
(1/2)r_1^2(\alpha_1-\sin \alpha_1)+(1/2)r_2^2(\alpha_2-\sin \alpha_2),
\eeq
where
\[
\cos (\alpha_1/2)=\frac{1}{2r_1}\left(r+\frac{r_1^2-r_2^2}{r}\right),
\]
\beq
\cos (\alpha_2/2)=\frac{1}{2r_2}\left(r-\frac{r_1^2-r_2^2}{r}\right).
\eeq
The function $\sigma(r)$ in (52) is just $\sigma(1,a,r)$. 

Formulas (49) -(54) allow to calculate numerically the average
$\langle S_n\rangle$ for any finite value of $N$ without much difficulty.

In the thermodynamic limit, $N\rightarrow\infty$ with $\eta $ being fixed,
 the calculation of $\langle S_n\rangle$ becomes trivial.
Indeed then one can neglect the part of integration in $r$ with $r>1-a$
altogether, with an error $\sim 1/a\sim1/\sqrt{N}$. With the same precision
one then finds
\beq
\langle S_n\rangle=S C_N^n(\sigma_0/S)^n(1-\sigma_0/S)^{N-n},
\eeq
where we have put $S_0\simeq S$
The physically relevant values of $n$ remain finite as 
$N\rightarrow\infty$. So we can approximately take
\beq
C_N^n=N^n/n!,\ \ \ (1-\sigma_0/S)^{N-n}=\exp(-N\sigma_0/S).
\eeq
We then find that in the thermodynamic limit the distribution of overlaps 
in $n$ is
Poissonian with the mean value given by $\eta$ (Eq. (13)).

Calculation of the multiplicity dispersion requires knowledge of the 
average of its
square. With the centers of the discs at $r_1,...r_N$ it has the form
\beq
\mu^2(r_1,...r_N)=(1/\sigma_0^2)(\sum_n\sqrt{n}S_n(r_1,...r_N))^2,
\eeq
where $S_N(r_1,...r_N)$ is given by (48). Taking the average over the 
discs centers
positions we now come to a double integral in $r$ and $r'$
\[
\langle\mu^2\rangle=\frac{1}{\sigma_0^2}\sum_{m,n}\sqrt{mn}
\int_S d^2rd^2r'\frac{1}{S_0^N}\int_{S_0}\prod_{i=1}^N d^2r_i\]\[
\sum_{\{i_1,..i_n\}\subset\{1,...N\}}\prod_{k=1}^{n}
\theta(a-|{\bf r}-{\bf r}_{i_k}|)
\prod_{k=n+1}^N\theta(|{\bf r}-{\bf r}_{i_k}|-a)\]\beq
\sum_{\{j_1,..j_m\}\subset\{1,...i\}}\prod_{l=1}^{m}
\theta(a-|{\bf r}-{\bf r}_{j_l}|)
\prod_{l=m+1}^N\theta(|{\bf r}-{\bf r}_{j_l}|-a).
\eeq

This complicated expression, however, continues to be factorized in all $r_i$.
 Let us assume for the moment that $n\geq m$. We can always
rename the variables $r_i$ to have in the first sum $r_1, ...r_n$ as  
variables in
the first product of $\theta$-functions
(and as a result, the rest $N-n$ variables $r_{n+1},..r_N$ go into the second
product). We shall then have $C_N^n$ terms with the identical first sum.
Now let $p$ variables in the
first product of $\theta$-functions in it coincide with $p$ variables from the
set $r_1,...r_n$. We have $C_n^p$ such terms, which will evidenly all have 
the same
dependence on the mentioned $p$ variables. The left $m-p$ variables from the
first product of $\theta$-functions in the second sum do not coincide with any
variables $r_1,...r_n$, so they are chosen fron variables $r_{n+1},...r_N$.
We shall have $C_{N-n}^{m-p}$ various terms of this sort. Thus the overall 
symmetry
factor turns out to be a multibynomial coefficient
\beq
C_N^nC_N^pC_{n-n}^{m-p}=\frac{N!}{p!(n-p)!(m-p)!(n-n-m+p)!}\equiv
C_N^{p,n-p,m-p}.
\eeq
It multiplies the result of integration over all $r_i$, $i=1,...N$, which 
has the form
(at fixed $r$ and $r'$)
\beq
\phi^p(r,r')\chi^{n-p}(r,r')\chi^{m-p}(r',r)\zeta^{N-n-m+p}(r,r'),
\eeq
where
\[
\phi(r,r')=(1/S_0)\int_{S_0}\theta(a-|{\bf r}-{\bf r}_1|)
\theta(a-|{\bf r'}-{\bf r}_1|),
\]
\[
\chi(r,r')=(1/S_0)\int_{S_0}\theta(a-|{\bf r}-{\bf r}_1|)
\theta(|{\bf r'}-{\bf r}_1|-a)=F(r)-\phi(r,r'),
\]
\beq
\zeta(r,r')=(1/S_0)\int_{S_0}\theta(|{\bf r}-{\bf r}_1|-a)
\theta(|{\bf r'}-{\bf r}_1|-a)=1-F(r)-F(r')+\phi(r,r'),
\eeq
with $F(r)$  defined before by (50).
Of course one should sum over all possible values of $p$.

Combining all the terms, we find the expression for the average square of
multiplicity as
\[
\langle\mu^2\rangle=\frac{1}{\sigma_0^2}\sum_{n,m,p}\sqrt{(n+p)(n+p)}\]\beq
C_N^{n,m,p}\int_Sd^2rd^2r'\phi^p(r,r')\chi^n(r,r')\chi^{m}(r',r)
\zeta^{N-n-m-p}(r,r').
\eeq

This expression is exact and may serve as a basis for the calculation of the
average square of the multiplicity at finite $N$. However the new function 
$\phi$  becomes very complicated when both variables $r$ and $r'$
are greater than $1-a$  (it is then given by the overlap area of three circles
and we do not know any simple analytic expression for it).

For this reason rather than analyze the general expression (62)  for 
finite $N$ we shall
immediately take the thermodynamic limit $N\rightarrow \infty$. We are in fact
interested in the dispersion, not in the average square of multiplicity. It is
important, since the leading terms in $N$ cancel in the dispersion. 
So we shall study
the difference
\beq
D^2=\langle\mu^2\rangle-\langle\mu\rangle^2
\eeq
in the limit $N\rightarrow\infty$, $\eta$ finite. As we shall see, 
although both terms
in the right-hand side of of (63) behave as $N^2$ separately, their 
difference grows
only as $N$.

Separating from (62) the term with $p=0$ and combinig it with the second term on
the right-hand side of (63) we present the total dispersion squared as a 
sum of two
terms
\beq
D^2=D_1^2+D_2^2,
\eeq
where
\[
D_1^2=
\frac{1}{\sigma_0^2}\sum_{n,m}\sqrt{nm}
\int_Sd^2rd^2r'[C_N^{n,m}
\chi^n(r,r')\chi^{m}(r',r)\zeta^{N-n-m}(r,r')-\]\beq
C_N^nC_N^mF^n(r)F^M(r')(1-F(r))^{N-n}(1-F(r'))^{N-m}]
\eeq
and $D_2^2$ is given by (62) with a restriction $p\geq 1$.

Function $S_0\phi(r,r')$ gives the overlapping area of three circles: 
the unit circle
$S_0$ and two circles $C$ and $C'$ of radii $a$ with centers at $r$ and $r'$.
It is evidently zero
if $R=|{\bf r}-{\bf r'}|>2a$ since for such $R$  circles $C$ and $C'$ do not 
overlap.
Because of this, in the part $D_2^2$, which has at least one factor $\phi$, the
integration in $r$ and $r'$ is restricted to the domain $R<2a$, whereas in the
part $D_1^2$ the integration covers the whole range of values of $R$.

We begin with the part $D_1^2$. Splitting the integration region in 
$r$ and $r'$ in 
two parts, $R>2a$ and $R<2a$ we make use of (61) and present the first part as
\[
D_{11}^2=
\frac{1}{\sigma_0^2}\sum_{n,m}\sqrt{nm}
\int_{R>2a}d^2rd^2r'F^n(r)F^m(r')\]
\beq[C_N^{n,m}(1-F(r)-F(r'))^{N-n-m}-
C_N^nC_N^m(1-F(r))^{N-n}(1-F(r'))^{N-m}].
\eeq
One notices immediately that the in the thermodynamic limit the leading terms
in the integrand (independent of $N$) cancel and only terms of the order $1/N$
remain. The integration over $r$ and $r'$ provides a factor $\propto N^2$ so
that the total contribution results $\propto N$.
To find this term we use that at large $N$, up to terms of the order $1/N^2$,
\[C_N^n=\frac{N^n}{n!}\left(1-\frac{n(n-1)}{2N}\right),\]
\[C_N^{m,n}=\frac{N^{n+m}}{n!m!}\left(1-\frac{(m+n)(m+n-1)}{2N}\right),\]
\[(1-F)^{N-n}=e^{-NF}(1+nF-NF^2/2),\]
where we have taken into account that $F$ has order $1/N$.
Then we obtain
\[
D_{11}^2=
\frac{1}{\sigma_0^2}\sum_{n,m}\frac{\sqrt{nm}}{n!m!}N^{n+m}
\int_{R>2a}d^2rd^2r'F^n(r)F^m(r')\exp(-N(F(r)+F(r'))\]\beq
(mF(r)+nF(r')-NF(r)F(r')-nm/N).
\eeq
The integrand  is now explicitly of the order $1/N$, so that we can  change the
integration region to $r,r'<1-a$, since the difference in the area will be of the
order $a^2\sim 1/N$, which results in the overall difference of the order $1/N^2$
and can safely be neglected. In the region $r,r'<1-a$ both $F(r)$ and $F(r)$
are constants, given by (51). Integration over $r$ and $r'$ gives an 
overall factor
$S^2$, so that in the end we get
\beq
D_{11}^2/N=\frac{1}{\eta^2}
e^{-2\eta}\sum_{n,m}\frac{\eta^{n+m}}{n!m!}\sqrt{nm}[\eta(n+m)-\eta^2-nm]=
-\frac{(\eta\langle\sqrt{n}\rangle-\langle n\sqrt{n}\rangle)^2}{\eta^2}.
\eeq
In the last expression the averages are to be taken over the Poissonian 
distribution
with the mean value $\eta$. The part $D_{11}^2$ is thus negative (and results
comparatively small for all values of $\eta$.

In the second part of $D_1^2$ the integration goes over a small region $R<2a$,
of the order $a^2\sim 1/N$, so that one can retain only the leading terms in the
integrand. Passing to the integration over $r$ and $R$ one observes that  at 
fixed
$r<1-a$ the integration region over $R$ covers the whole region $R<2a$. 
For $r>1-a$
the integration region in $R$ becomes much more complicated, determined by the
condition $r'=|{\bf r}+{\bf R}|<1+a$. However the contribution from the 
latter region
will be of the order $a^3\sim1/N\sqrt{N}$, since 
apart from a factor $\propto a^2$ from the integration over $R$, a factor 
$\propto a$
appears due to integration over $r>1-a$. 
So, up to terms of the relative order $1/\sqrt{N}$, we can neglect
the contribution from the region $r>1-a$.

If $r<1-a$ the circle $C$ is completely inside the unit circle $S_0$. Then its
intersection with the circle $C'$ is also automatically inside $S_0$. Therefore
function $S_0\phi(r,r)$ in this region is simply given by the overlap of 
the circles
$C$ and $C'$, that is $\sigma(a,a,R)$ defined by (53). We thus find
\beq
\phi(r,r')=\frac{\sigma_0}{S}\lambda(R),
\eeq
where
\beq
\lambda(R)=(1/\pi)(\alpha-\sin\alpha),\ \ \alpha=2\arccos(R/2).
\eeq
Putting this into the expression for $D_{12}^2$ and retaining the leading 
terms in the
limit $N\rightarrow\infty$ we obtain
\beq
D_{12}^2/N=\frac{2}{\eta}e^{-2\eta}\sum_{n,m}\frac{\sqrt{nm}}{n!m!}\eta^{n+m}
\int_0^2RdR[(1-\lambda(R))^{n+m}e^{\eta\lambda(R)}-1].
\eeq
This expression can be easily evaluated numerically. It is relatively 
large and also
negative.

We finally come to the part $D_2^2$. The integration in $r,r'$ goes over 
$R<2a$, so
that we can apply the same approximations as made in calculating $D_{12}^2$. 
We find
\beq
D_{2}^2/N=\frac{2}{\eta}e^{-2\eta}\sum_{n,m}\sum_{p=1}
\frac{\sqrt{(n+p)(m+p)}}{n!m!p!}\eta^{n+m+p}
\int_0^2RdR\lambda^p(R)(1-\lambda(R))^{n+m}e^{\eta\lambda(R)}.
\eeq 
This part is evidently positive. Its numerical evaluation shows that it 
nearly cancels
the large negative contributions from $D_1^2$ (in fact four digits are cancelled
typically). So the numerical calculation of the dispersion requires some care.

\section{References}

\hspace{0.6cm}[1]  A.Capella, U.P.Sukhatme, C.-I.Tan and J.Tran Thanh
Van, Phys. Lett. {\bf B81}  68 (1979); Phys. Rep. {\bf 236}  225 (1994).

[2] A.B.Kaidalov and K.A.Ter-Martirosyan, Phys. Lett. {\bf B117} 247  (1982).

[3] B.Andersson, G.Gustafson and B.Nilsson--Almqvist, Nucl.
Phys. {\bf B281}  289 (1987).

[4] K.Werner, Phys. Rep. {\bf 232}  87 (1993).

[5] M.Gyulassy, CERN preprint CERN -TH 4794 (1987)

[6] H.Sorge, H.Stoecker and W.Greiner, Nucl. Phys. {\bf A498}  567c (1989).

[7] N.Armesto, M.A.Braun, E.G.Ferreiro and C.Pajares,
Phys. Rev.Lett. {\bf 77} 3736  (1996).

[8] M.A.Braun, C.Pajares and J.Ranft, Santiago de Compostela
preprint US-FT/24-97, hep-ph/9707363.

[9] M.Nardi and H.Satz, Bielefeld preprint BI-TP 98/10, hep-ph/9805297;
H.Satz, Bielefeld preprint BI-TP 98/11, hep-ph/9805418.

[10] T.S.Biro, H.B.Nielsen and J.Knoll, Nucl. Phys. {\bf B245} 449  (1984).

[11] M.A.Braun and C.Pajares, Phys. Lett. {\bf B287}  154 (1992);
Nucl. Phys. {\bf B390} 542,549  (1993).

[12] N.S.Amelin, M.A.Braun and C.Pajares, Phys. Lett. {\bf B306} 312 (1993);
Z.Phys. {\bf C63}  507 (1994).

[13] A.Bialas and W.Czyz, Nucl. Phys. {\bf B267} 242  (1986) .

[14] D.Stauffer, Phys. Rep. {\bf 54} 2  (1979).

[15] M.B.Isichenko, Rev. Mod. Phys.{\bf 64}  961 (1992).

[16] J.D.Bjorken, Phys.Rev {\bf D 27} (1983) 140. 

[17] See e.g. A.Capella, A.Kaidalov and J.Tran Thanh Van,
Orsay preprint LPT ORSAY 99-15, hep-ph/9903244.

\section{Figure captions}

\hspace{0.6cm}Fig. 1. Projections of two overlapping strings onto the transverse plane.

Fig. 2. Damping of the multiplicity as a function of $\eta$.

Fig. 3. Percolation dispersion squared of the multiplicity  per string
in units $\mu_0^2$ as a function of $\eta$.

Fig. 4. Fraction of the total interaction area covered by the maximal cluster as
a function of $\eta$.

Fig. 5. Parameter $b$ of the FBC  for different rapidity intervals
as a function of $\eta$ for fixed $N$.

\section{Table captions}
Columns show from left to right:
c.m. energy per nucleon $\sqrt{s}$; parameter $\eta$ (Eq. (1));
number of initially produced strings $N$; central charged multiplicity per unit
rapidity $\mu$ and its FBC parameter $B(\mu)$, the FBC parameter for the
inverse $\avpt$, $b(\tau)$, and the multiplicities and $b(\mu)$ in the
independent string model (subscript 0).

\newpage

\section{Table}

\begin{center}
{\bf S-S scattering ($b=0$)}\vspace{0.6cm}

\begin{tabular}{|r|r|r||r|r|r||r|r|r|}\hline 
$\sqrt{s}$&$\eta$&$N$&$\mu$&$b(\mu)$&$b(\tau)$&$\mu_0$&$b_0$
\\\hline
19.4&0.33& 99& 101&0.67& 0.04& 111& 0.65\\\hline
62.5&0.47& 143& 140&0.74& 0.08& 160& 0.72\\\hline
 200&0.65& 198& 185&0.79& 0.13& 221& 0.77\\\hline
 546&0.84& 255& 228&0.82& 0.18& 284& 0.80\\\hline
1800&1.10& 336& 283&0.85& 0.24& 374& 0.82\\\hline
7000&1.47& 448& 349&0.87& 0.30& 499& 0.85\\\hline
\end{tabular}
\end{center}
\vspace*{0.6 cm}

\begin{center}
{\bf Pb-Pb scattering ($b=0$)}\vspace{0.8cm}

\begin{tabular}{|r|r|r||r|r|r||r|r|}\hline 
$\sqrt{s}$&$\eta$&$N$&$\mu$&$ b(\mu)$&$b(\tau)$&$\mu_0$&$b_0$
\\\hline 
  19.4&1.45 &1530&$1200$&$0.69$&0.13&$ 1700$&$ 0.65$\\\hline
  62.5&2.09 &2200&$1530$&$0.75$&0.19&$ 2450 $&$ 0.72$\\\hline
 200.0&2.88&3040&$1870$&$0.78$&0.25&$  3390$&$ 0.77$\\\hline
 546.0&3.72&3920&$2170$&$0.80$&0.28&$ 4370$&$ 0.80$\\\hline
1800.0&4.90&5170&$2530$&$0.81$&0.31&$ 5760$&$ 0.82$\\\hline
7000.0&6.53&6890&$2940$&$0.82$&0.33&$ 7680$&$ 0.85$\\\hline
\end{tabular}
\end{center}

\newpage

\section{Figures}
\textwidth 210 mm
\begin{center}

\begin{figure}[htb]
\hspace{1.5cm}\epsfig{file=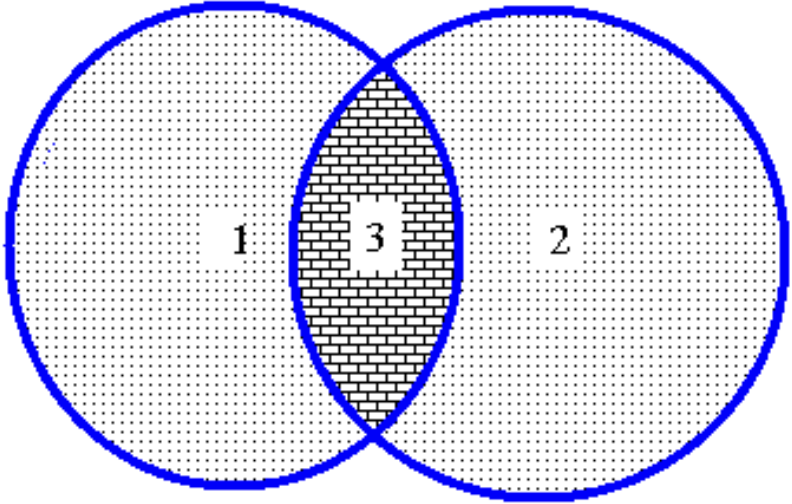,bbllx=40pt,bblly=0pt,bburx=680pt,bbury=500pt,%
width=20cm}
\caption{ Projections of two overlapping strings onto the transverse plane.}
\label{Fig.1}
\end{figure}

\begin{figure}[htb]
\hspace{1.5cm}\epsfig{file=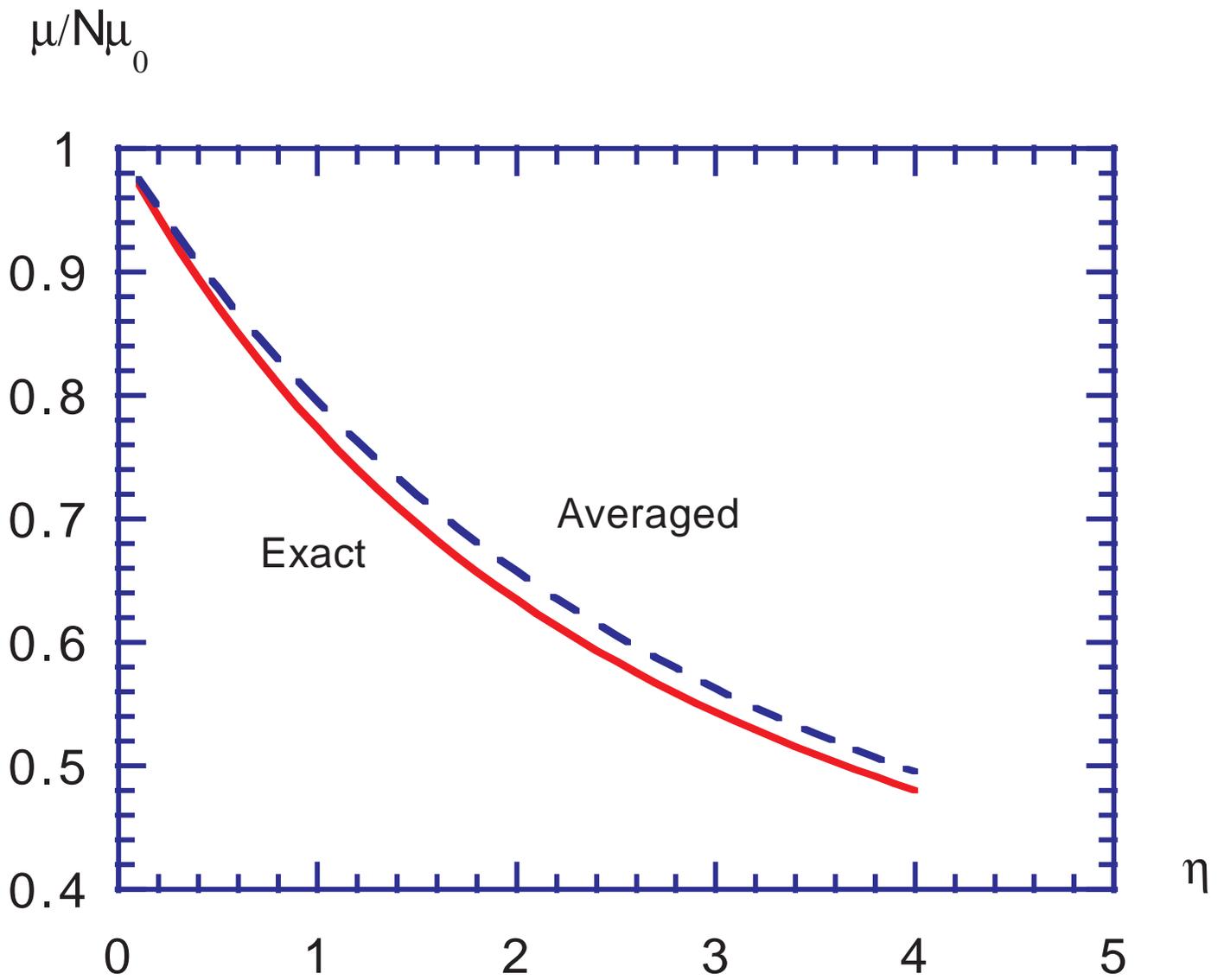,bbllx=184pt,bblly=258pt,bburx=530pt,bbury=532pt,%
width=20cm}
\caption{ Damping of the multiplicity as a function of $\eta$.}
\label{Fig.2}
\end{figure}

\begin{figure}[htb]
\hspace{1.5cm}\epsfig{file=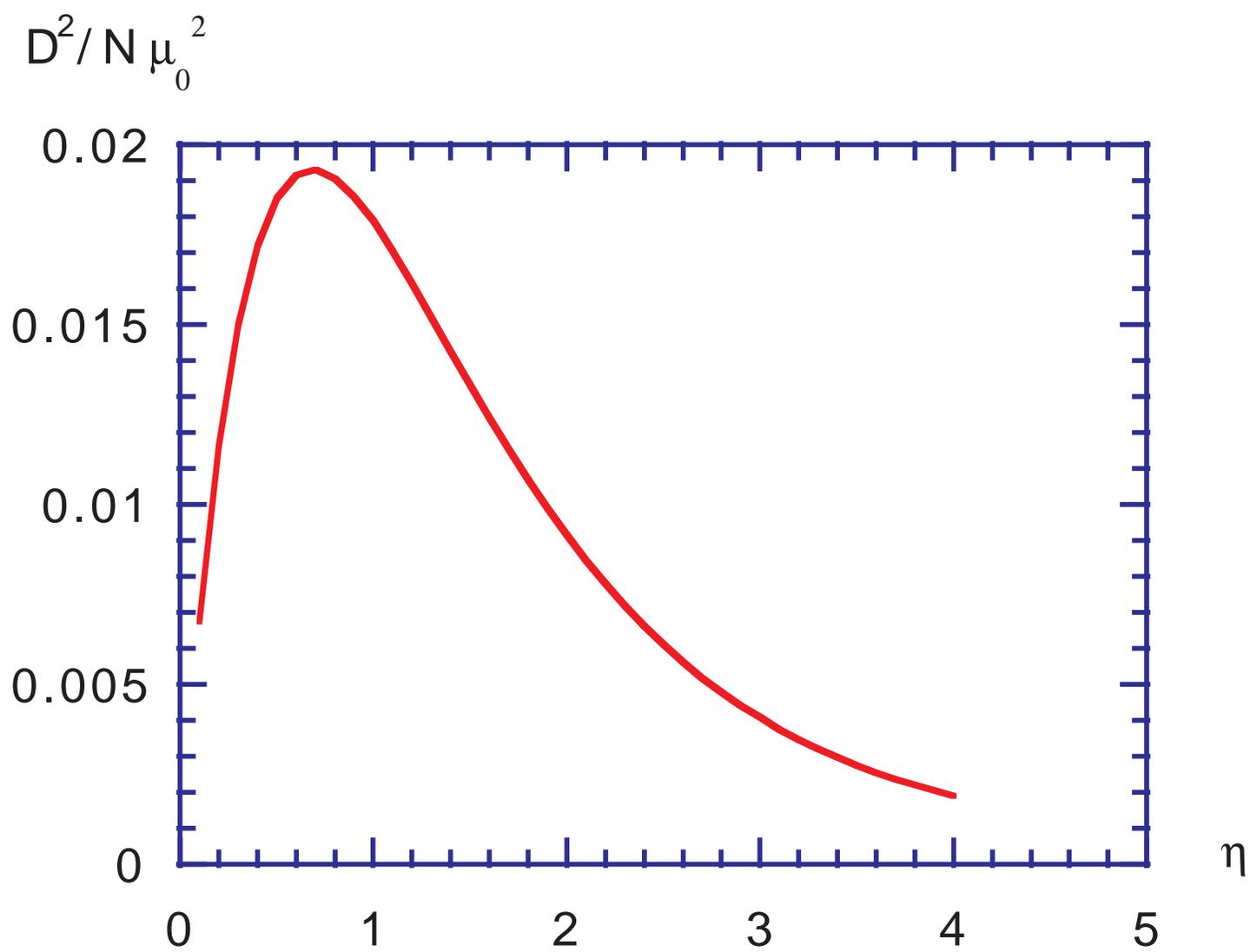,bbllx=171pt,bblly=259pt,bburx=529pt,bbury=526pt,%
width=20cm}
\caption{Percolation dispersion squared of the multiplicity  per string
in units $\mu_0^2$ as a function of $\eta$.}
\label{Fig.3}
\end{figure}

\begin{figure}[htb]
\hspace{1.5cm}\epsfig{file=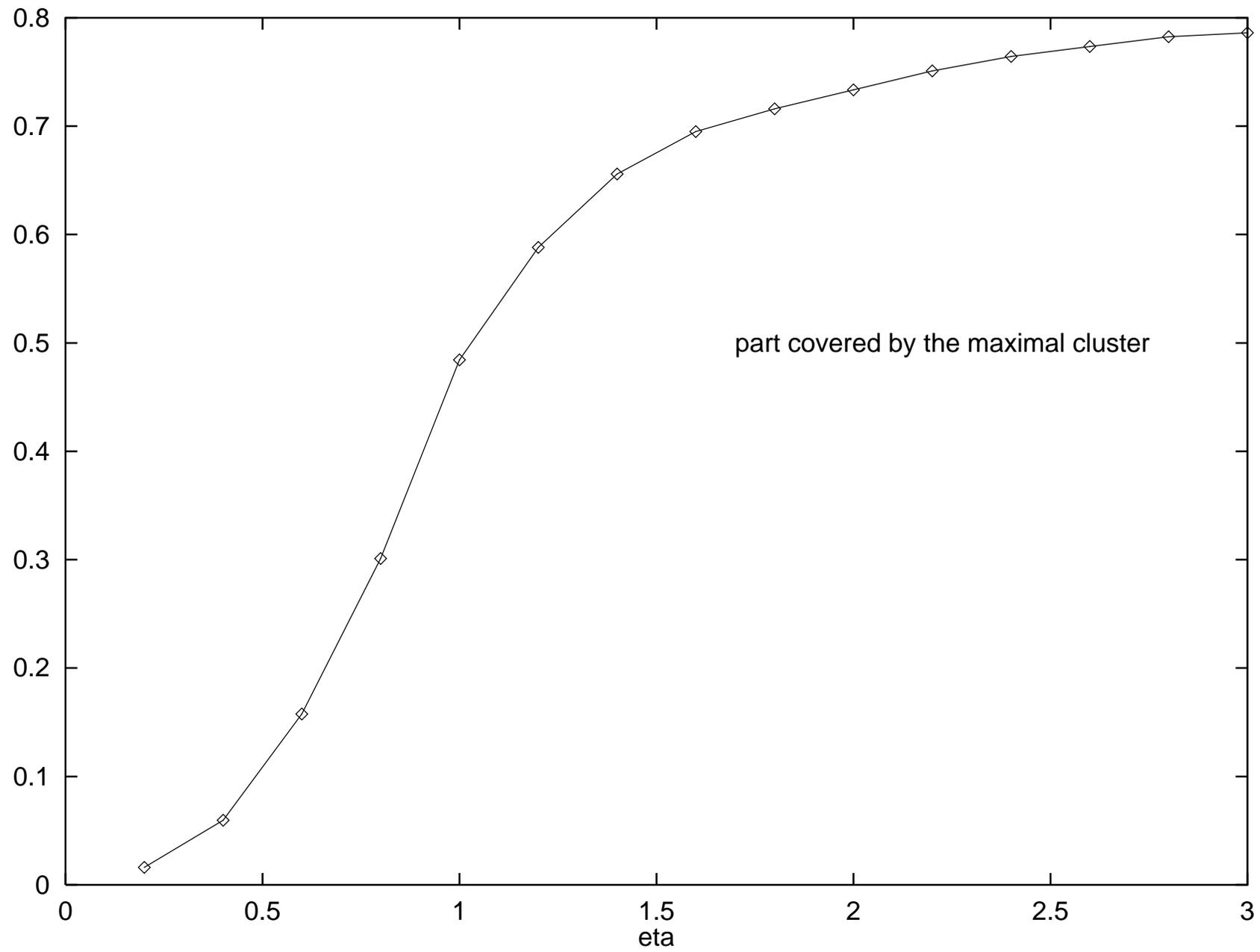,bbllx=40pt,bblly=69pt,bburx=640pt,bbury=785pt,%
width=20cm}
\caption{Fraction of the total interaction area covered by the maximal 
cluster as a function of $\eta$.}
\label{Fig.4}
\end{figure}

\begin{figure}[htb]
\hspace{1.5cm}\epsfig{file=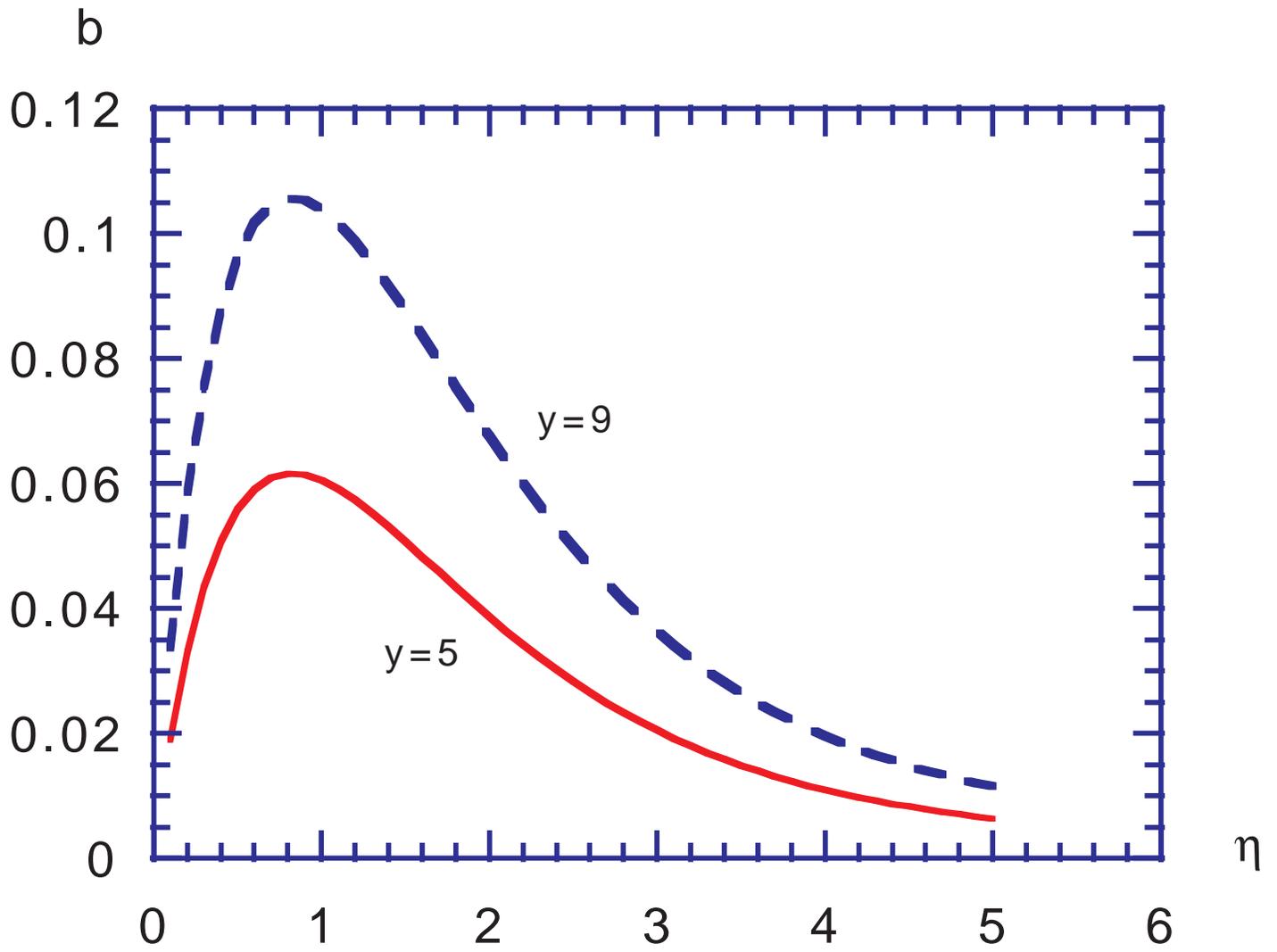,bbllx=178pt,bblly=259pt,bburx=531pt,bbury=521pt,%
width=20cm}
\caption{Parameter $b$ of the FBC  for different rapidity intervals
as a function of $\eta$ for fixed $N$.}
\label{Fig.5}
\end{figure}
\end{center}

\end{document}